\documentclass[aps,prl,twocolumn,groupedaddress]{revtex4}
\usepackage{epsf}
\usepackage{graphicx}

\newcommand{\strelki}{\mathord{\buildrel{\lower3pt\hbox{$\scriptscriptstyle\leftrightarrow$}}
\over g}}
\newcommand{\ustrelki}{\mathord{\buildrel{\lower3pt\hbox{$\scriptscriptstyle\leftrightarrow$}}
\over g_{u}}}
\newcommand{\sstrelki}{\mathord{\buildrel{\lower3pt\hbox{$\scriptscriptstyle\leftrightarrow$}}
\over g_{s}}}

\begin{document}

\title{ESR of coupled spin-1/2  chains: unveiling zig-zag-type interchain correlations}
\author{A. A. Validov$^1$,  M. Ozerov$^2$, J. Wosnitza$^2$,  S. A. Zvyagin$^2$,  \\
M. M. Turnbull$^3$, C. P. Landee$^3$ and G. B. Teitel'baum$^1$}\email{grteit@kfti.knc.ru}
\affiliation{$^1$E.K. Zavoiskii Institute for Technical Physics of the RAS, Kazan 420029, Russia}
\affiliation{$^2$Dresden High Magnetic Field Laboratory (HLD),Helmholtz-Zentrum Dresden-Rossendorf,
01328 Dresden, Germany}
\affiliation{$^3$Clark University, Worcester, MA 01610, USA}
\date{\today}

\begin{abstract}
By means of electron spin resonance investigations we revealed the crucial role of the
interchain coupling in the spin dynamics of the spin-1/2 Heisenberg antiferromagnetic (AF) chain
material copper-pyrazine-dinitrate, Cu(C$_4$H$_4$N$_2$)(NO$_3$)$_2$. We found that the dominating
interchain interaction is of a zig-zag type. This interaction gives rise to geometrical frustration
effects and strongly influences the character of AF ordering. Combining our experimental findings with
the results of a quasiclassical approach we argue that at low temperatures the system orders in an
incommensurate spiral state.

\end{abstract}

 \maketitle



\bigskip
Recently, a large amount of theoretical and experimental work has been addressed to low-dimensional
(low-D) spin systems.  Spin-1/2 AF chains can be regarded as one of the most important
quantum-mechanical model systems, exhibiting a rich variety of unusual magnetic properties
\cite{Sachdev}. The excitation spectrum of an ideal isotropic  spin-1/2 AF chain is gapless and
dominated by a so-called \textquotedblleft spinon continuum\textquotedblright \cite{Muller}. Since the
spin-1/2 AF chain is critical, even small perturbations can significantly change fundamental
properties of the system. One of the most prominent examples is a perturbation by an alternating $g$
tensor or by the Dzyaloshinskii-Moriya interaction \cite{OA2002}.  The magnetic excitation spectrum in
such a system is formed by solitons and their bound states (breathers), and the system exhibits a
field-induced gap \cite{Zvyagin_1}. Interchain coupling can be regarded as another type of
perturbation, naturally present in all low-D materials.  It can be considered as a tuning parameter
competing with other perturbations  and determining quantum critical properties of spin systems at low
temperatures. In addition, the interchain coupling can add frustration aspects to the problem due to a
possible zig-zag-type interaction \cite{White, Starykh, White1}, typical also for spin systems on a
triangular lattice. The understanding of all these issues is crucial for elucidating the
low-temperature dynamics of quantum magnets.  In addition to a comprehensive theoretical analysis,
progress on the experimental side provides growing opportunities to study quantum critical phenomena
in low-D spin systems.

The AF chain compound  copper pyrazine dinitrate (Cu(C$_4$H$_4$N$_2$)(NO$_3$)$_2$, abbreviated as
CuPzN) \cite{Santoro} is regarded as the best realization of an isotropic quantum spin-1/2 Heisenberg
system  \cite{Hammar,Stone}. CuPzN has  an exchange coupling along the chains of $J/k_B=10.3$ K,
showing no long-range order down to $\sim 0.1$ K \cite{Lancaster}. The latter indicates that a
perturbation term $J'$, eligible for a possible violation of the spin symmetry, is extraordinarily
small. It makes CuPzN an ideal object to investigate even  tiny perturbation effects in Heisenberg AF
spin-1/2 chains.

Here, we present results of electron spin resonance (ESR) studies of CuPzN, allowing us to probe the quantum critical behavior
of coupled spin chains in the vicinity of the critical point, separating the Luttinger-liquid and the
magnetically ordered phases. ESR appears to be an effective tool to answer to the question crucial for
understanding the critical behavior: which type of interchain interaction is more effective - the
transverse or zig-zag coupling? The former gives rise to a staggered field and a field-induced
spin-gap, while for the latter the effective field transferred from the adjacent chains is cancelled
and spin frustration is expected. Analyzing the specific ESR properties we studied the influence of
possible geometrically frustrating interactions in CuPzN and proposed a microscopic picture of the
magnetic ordering in this compound.

\begin{figure}[tbp]
\centering \includegraphics[height=6cm]{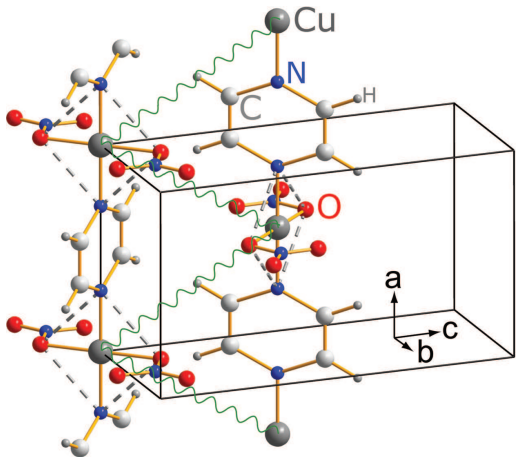} 
\caption{(Color online) Schematic structure of CuPzN. The canting of the basal planes (dashed
lines) for nonequivalent Cu sites is shown enlarged for the sake of clarity. The intrachain coupling,
$J$, is along the \textbf{a} axis. The possible interchain couplings are either between the equivalent
Cu sites along \textbf{b} (transverse coupling)  or connecting nonequivalent Cu in a zig-zag manner
as shown by the wavy lines.}
\end{figure}

Plate-like single crystals of CuPzN with typical sizes of 2x1x0.1 mm$^3$ were used in our experiments.
The samples were grown  by evaporation of a mixture of pyrazine with an aqueous solution of copper
nitrate \cite{Hammar}. X-band ESR experiments were performed at the Zavoiskii Institute for Technical
Physics of the Russian Academy of Sciences by use of the spectrometer Bruker BER 318. High-field ESR
experiments were performed at the Dresden High Magnetic Field Laboratory (Hochfeld-Magnetlabor
Dresden) employing a~multiple-frequency ESR spectrometer (similar to that described in
\cite{spectrometer}) equipped with a 16 T superconducting magnet and VDI  sources of millimeter-wave
radiation (product of Virginia Diodes Inc.).

The crystal structure of CuPzN is schematically shown in Fig. 1. It belongs to the orthorhombic space
group $Pnma$ with two formula units per unit cell. The cell parameters measured at room temperature
\cite{Santoro} show very slight contraction upon cooling down to 2 K  \cite{Jornet}: $a$ - from 6.712
\AA\ to 6.69166(6) \AA;  $b$ - from 5.142 \AA\ to 5.10538(4) \AA ; $c$ - from 11.732 \AA\ to
11.60022(9) \AA. There are two nonequivalent spin chains running along the $a$ axis. The
CuN$_{2}$O$_{2}$ basal planes for two Cu-sites from neighboring chains are canted by $\pm 5.2^{\circ
}$ (measured at 4 K) from the crystallographic $ac$ plane. Cu-sites from neighboring chains are
shifted by half a period along $a$ in the $ac$ planes, which is not the case for the chains in the
$ab$ planes. It is reasonable to expect that the quantum critical properties will be determined by an
interchain coupling which is more effective, i.e., either by the transverse spin-exchange coupling
between Cu-sites from the unshifted chains in the $ab$ planes, or by a zig-zag-type exchange coupling
between Cu ions belonging to the chains in the  $ac$ planes. In the first case, the interchain
interaction gives rise to a staggered field and spin-gap, in the second case the effective field
transferred from adjacent chains cancels and spin frustration is expected.
\begin{figure}[tbp]
\centering \includegraphics[height=5cm]{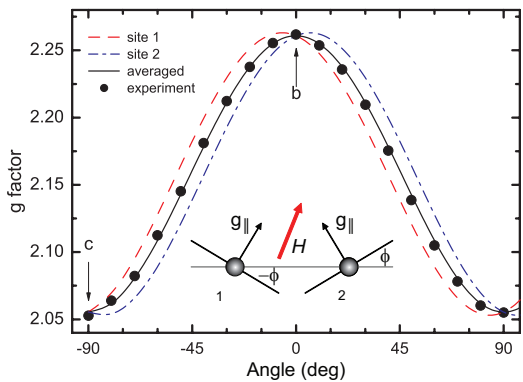}
\caption{(Color online) Angular dependence of the $g$-factor with magnetic field applied in the $bc$
plane at 4.2 K (circles). The solid black line is the result of the superposition of two angular
dependences (shown by the dashed and dash-dotted lines) originated from two inequivalent Cu sites.}
\end{figure}

ESR experiments performed at 9.4 GHz and at temperatures down to 1.4 K  revealed a single line with a
shape very close to Lorentzian; only below 10 K were deviations from such a shape found. The almost
temperature-independent $g$-factors, $g_{a}$ =2.053, $g_{b}$  = 2.265, and $g_{c}$ = 2.063, are in
good agreement with previously published ESR data \cite{McGregor}. Figure 2 shows the angular
dependence of the $g$-factor measured at 4.2 K (symbols). Assuming tetragonal symmetry, we can
describe the experimentally obtained
 angular dependence taking into account contributions from two inequivalent Cu-sites (shown by the red
 dashed and blue dash-dotted lines). Note that each site has the  same $g_{\|}$ and $g_{\bot}$, in the corresponding
coordinate  system tilted by $\phi=\pm5.2^{\circ})$ from the basal plane (a schematic view along the
$a$ axis is shown in the inset of Fig. 2). As a result, the $g$-tensor for inequivalent sites can be
described  by
\begin{equation}\label{Eq1}
    \strelki=\left(
\begin{array}{ccc}
2.0526 & 0 & 0 \\
0 & 2.0543 & \pm 0.0189 \\
0 & \pm 0.0189 & 2.2607%
\end{array}%
\right),
\end{equation}
with alternating signs for neighboring chains.

At temperatures above 10 K, the observed linewidth is relatively small ($\Delta H\sim 20-30$ Oe) and
shows no indication of line broadening due to spin diffusion.
\begin{figure}[tbp]
\centering \includegraphics[height=9cm]{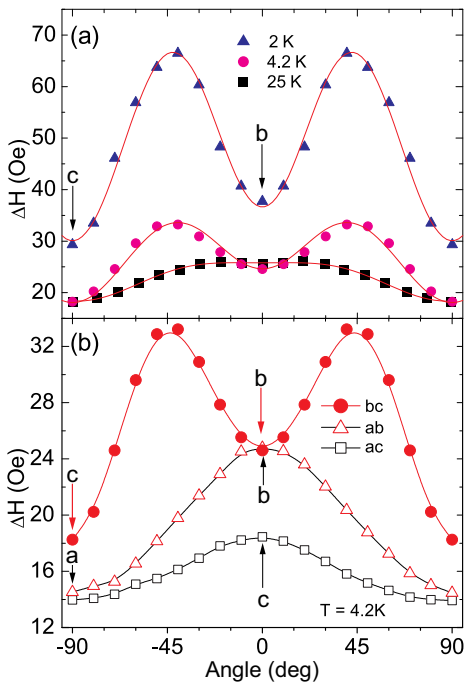} 
\caption{(Color online) (a) Angular dependence of the ESR linewidth  obtained at 9.4 GHz and at
temperatures of 25 K (squares), 4.2 K (circles), and 2 K (triangles); the magnetic field is in the
$bc$ plane.  b) Angular dependence of the ESR linewidth obtained at 9.4 GHz and at 4.2 K for magnetic
fields in the $bc$ plane (circles), $ab$ plane (triangles), and $ac$ plane (squares). Lines correspond
to fit results (see the text for details).}
\end{figure}
On the other hand, at lower temperatures,  $T \lesssim J/k_{B}$, drastic angular-dependent changes in
the linewidth, $\Delta H$, (with field applied in the $bc$ plane)  can be observed (Fig. 3). Two
maxima located  near $\pm 45^{\circ}$ appear. (Note, that such a behaviour for magnetic fields applied
in the $ac$ or $ab$ plane does not occur). Interestingly, the ESR linewidth exhibits a strong increase
at low temperatures when the magnetic field is applied in the directions corresponding to these maxima
 (Fig. 4).

To explain the origin of such dramatic  changes let us turn to  results of high-frequency ESR
measurements (Fig. 5). The experiments performed at about 55.2, 110.4, 220.8, and 331.2 GHz allowed us
to clearly resolve two ESR lines with $g_1 =  $2.13(6) and $g_2  =  $2.17(4) for  magnetic fields
applied at $45^{\circ}$ with respect to the $b$ axis; that  corresponds to two inequivalent  Cu-sites.
Note that the $g$-factor values are consistent to those extracted from the X-band ESR data (Fig. 2),
for the same field orientation where the maximal line broadening was observed. A typical angular
dependence of the high-frequency ESR lineshape is depicted in the  inset (a) of Fig. 5 which shows ESR
spectra obtained at 1.4 K and 220.8 GHz with different orientations of the magnetic field in $bc$
plane. The ESR absorption is split if the magnetic field is applied at $45^{\circ}$ with respect to
the $b$ or $c$ directions (hereafter called $l$ direction), while no splitting is observed when the
magnetic field is applied along the $b$ and $c$ directions. The spectacular temperature evolution of
the signal is displayed in the inset (b) of Fig. 5 where ESR spectra at 1.4 and 42 K for the $l$
direction are shown. Two well-resolved ESR lines were observed at 1.4 K, which is contrary to the
signal at 42 K. Such a behaviour reflects a crossover between two different ESR regimes which is
driven by the temperature dependence of the interchain-hopping rate $\omega_{h}$. At higher
temperatures when the hopping rate exceeds the mismatch of the Zeeman frequencies for two
nonequivalent Cu-sites ($|(g_{1}-g_{2})|\beta H/\omega_{h} \ll 1$; this regime is known as
fast-fluctuation or line-narrowing regime) only one line is observed. Here, $\beta =\mu_{B}/\hbar$,
where $\mu_{B}$ is the Bohr magneton. Upon lowering the temperature, the spin system enters the
slow-fluctuation regime ($|(g_{1}-g_{2})|\beta H/\omega_{h} \gg 1$) and the two different
contributions from two nonequivalent sites can be resolved.
\begin{figure}[tbp]
\centering \includegraphics[height=5cm]{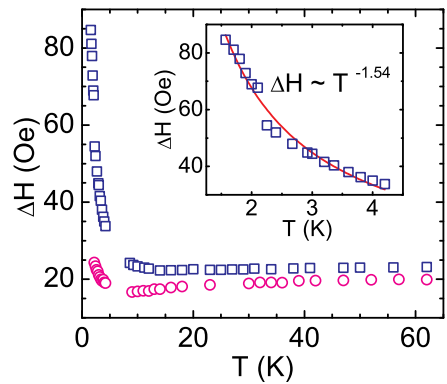} 
\caption{(Color online) Temperature dependence of the linewidth, $\Delta H$,  measured at 9.4 GHz with
magnetic fields applied along the $b$ direction (circles) and tilted by $45^{\circ}$ towards the $c$
axis (squares). Inset: Low-temperature part of the ESR linewidth for the $45^{\circ}$ orientation; the
line corresponds to a fit result (see the text for details).}
\end{figure}

Thus, the temperature evolution of the ESR signal, shown in the inset (b) of Fig. 5, gives the
unambiguous evidence of  the importance of the zig-zag-type exchange interaction (schematically shown
by wavy lines in Fig. 1) These observations are consistent with results of inelastic
neutron-scattering data \cite{Hammar}, revealing the absence of  substantial spin correlations along
the $b$ direction. As a result, the effective magnetic correlations of CuPzN can be described by an
anisotropic Heisenberg model on a triangular lattice (in the $ac$ plane) with the exchange along the
$a$ axis (intrachain exchange) equal to $J$ and the weak interchain zig-zag coupling $J_1$. It was
shown recently \cite{White1} that the main features of this model in the limit of small $J_1/J$ may be
captured with its simplest cartoon - the zig-zag two-leg spin ladder.

\begin{figure}[tbp]
\centering \includegraphics[height=7cm]{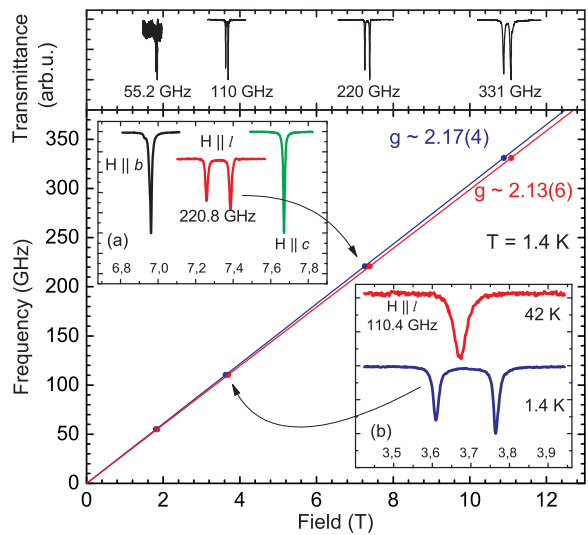} 
\caption{(Color online) Upper panel: High-frequency ESR spectra with magnetic fields applied along the
$l$ direction. Lower panel: Frequency-field dependence of the corresponding ESR absorptions. Inset (a)
shows ESR spectra at $T=1.4$ K and 220.8 GHz with magnetic field applied along $b$, $c$, and  $l$.
Inset (b) shows ESR spectra obtained at 1.4 and 42 K at 110.4 GHz with magnetic field applied along
$l$.}
\end{figure}

In the case of a zig-zag spin ladder the spin-Hamiltonian can be written as \cite{White}

\begin{equation}\label{Eq2}
  H = \sum_i [  J(\vec{S}_i \vec{S}_{i+1} +\vec{T}_i \vec{T}_{i+1}) +
      J_1\vec{S}_i ( \vec{T}_i +\vec{T}_{i+1} ) ],
\end{equation}
where $J$ and $J_1$ are intrachain and interchain exchange interactions,
correspondingly, while $S_i$  and $T_i$  are spin operators for  two adjacent crystallographic chains.

The magnitude of the interaction integral $J_1$ may be estimated from the analysis of the specific
linewidths' dependencies at 9.4 GHz (Fig. 3a),  which provides one more evidence of the triangular
interchain exchange. The linewidth,  $\Delta H$, may be written as a superposition of two
contributions
\begin{equation}\label{Eq3}
\Delta H=\Delta H_{chain} +(1/4) (g_{1}-g_{2})^{2}\beta^2 H^{2}/\omega_{h} ,
\end{equation}
where \ the first one is due to the anisotropy of the intrachain exchange, the dipolar coupling, etc.
\cite {OA2002}. The second term describes the inhomogeneous contribution which depends on the hopping
between the inequivalent sites of the adjacent chains.  This expression follows from the
fast-fluctuation limit of the imaginary part of the dynamical susceptibility at the frequency $
(g_{1}+g_{2})\beta H/2$ \cite{Slichter} and has the well-known form of an exchange-narrowed linewidth
\cite{Anderson}. Note also that the first term has a $180^{\circ}$ symmetry with maxima for magnetic
fields aligned along $b$ while the second one, due to the $(g_{1}-g_{2})^{2}$ factor, has a
$90^{\circ}$ symmetry with maxima close to the $45^{\circ}$ field orientation. We assume that the
hopping rate ($\omega_{h} $) for $T\gtrsim 10$ K becomes temperature independent and tends to
$4J_1/\hbar$ while for lower $T$  it may be expressed via the dynamic scaling relation
$\omega_{h}\approx \tau^{z\nu}$ \cite {Hoehenberg}. Here, $\tau$ is the normalized temperature, $z$ is
the dynamical critical exponent and $\nu$ is the critical exponent of the correlation length.

To explain  the origin of the unusual angular dependence of the ESR linewidth (Fig. 3)  the
experimental data were described using Eq. (3). For that, $g_1$ and $g_2$ obtained from the angular
dependence at $T=4.2$ K (Fig. 2) were used.  Note,  that since at high $T$ the hopping frequency is
temperature independent and equal to $4J_1/\hbar$ the fit to the  angular dependence of the linewidth
at $T=25$ K gives us $J_1/k_B$= 0.044 K. Fitting curves at this and at lower temperatures, where the
exchange-narrowing contributions are much more pronounced due to the slowing down of $\omega_{h}$ with
$z\nu \approx 1.54$, are shown in Fig. 3 by lines. Thus, the additional maxima  in the angular
dependence of the ESR linewidth
 observed at low temperatures ($T=4.2$ and $2$ K, Fig. 3) correspond to  specific
directions, where  the system approaches the boundary separating two types of spin-fluctuation regimes
($|(g_{1}-g_{2})|\beta H/\omega_{h} \sim 1$) and  the exchange narrowing process is suppressed. This
proves that the critical spin dynamic in  CuPzN is essentially determined by zig-zag-type interchain
correlations in the $ac$ plane.

Let us now turn to the discussion of the possible magnetic ordering of CuPzN, which should strongly be
affected by the zig-zag-type of spin interactions. As mentioned, CuPzN can be described using the
effective spin Hamiltonian  given by Eq. (2). Note, that this model is equivalent to a 1D spin-1/2
Heisenberg AF system, with nearest-neighbor (NN) and frustrating next-nearest-neighbor (NNN) couplings
($J_1$ and $J$, correspondingly) \cite{White}. A classical solution for this 1D chain system with NN
and NNN exchange interactions corresponds to an incommensurate (IC) spiral structure with pitch angle
$\theta$ \cite{White} given by
\begin{equation}\label{Eq4}
\cos \theta = - J_1/4J.
\end{equation}
For $J/k_B= 10.3$ K and $J_1/k_B = 0.044$ K the pitch angle is close to $\pi$/2:  $\theta = \pi/2$ +
0.0042. The correlation length $\xi$, which is given by the relation
\begin{equation}\label{Eq5}
\theta - \pi/2 \sim 1/\xi,
\end{equation}
may be estimated as $\xi\sim 240$ in the units of the lattice spacing $a$. Thus, because of the
frustrated nature of the interchain interactions the ground state in CuPzN corresponds to an IC AF
order, while a homogeneous AF ordering is impossible at any temperature \cite{White, White1}.

Using the semi-empirical expression for the N\'{e}el temperature \cite{Takayama} for a system of
coupled spin chains,
\begin{equation}\label{Eq6}
T_N=4cJ_1\sqrt{\ln(\lambda J/T_N)+(1/2)\ln\ln(\lambda J/T_N)},
\end{equation}
with $c$ = 0.233 and $\lambda$ = 2.6 for  $J_1/k_B = 0.044$ results in $T_N$ = 103 mK. It is worth
noting that this value is in good agreement with the ordering temperature (107 mK) found for CuPzN by
the use of zero-field muon-spin relaxation measurements \cite{Lancaster}. This indicates that the
magnetic ordering observed in \cite{Lancaster} corresponds to a transition to the IC spiral state
discussed above.

In conclusion, quantum critical properties of the quasi-1D Heisenberg AF system CuPzN have been probed
by means of ESR. The specific angular and temperature dependencies of the ESR absorption revealed the
important role of zig-zag-type interchain  exchange interactions. Using a quasiclassical description
and the obtained ESR data,  CuPzN is predicted to have an IC spiral ordered ground state at low
temperatures. According to the classification discussed in \cite{White} the smallness of the ratio
$J_{1}/J= 0.0043$ found in our work  makes it possible to conclude that CuPzN  belongs to an
interesting class of spin systems which are closely related to the generalized Kondo lattice.

\begin{acknowledgements}

The work of A.A.V. and G.B.T. was supported through the  RFBR grant N 10-02-01056. This work was
partly supported by the DFG and EuroMagNET (EU contract No. 228043).

\end{acknowledgements}

\bigskip

\begin{references}

\bibitem{Sachdev} S. Sachdev,  Quantum Phase Transitions (Cambridge: Cambridge University
Press) 1999.

\bibitem{Muller} G. M\"{u}ller, H. Thomas, H. Beck, and J.C. Bonner, Phys. Rev. B {\bf 24}, 1429 (1981).

\bibitem{OA2002} M. Oshikawa and I. Affleck,  Phys. Rev. B {\bf 65},
134410 (2002).


\bibitem{Zvyagin_1} S.A. Zvyagin, A.K. Kolezhuk, J. Krzystek, and R. Feyerherm,
Phys. Rev. Lett. {\bf 93}, 027201 (2004).

\bibitem{White} S.R. White  and I. Affleck,  Phys. Rev. B {\bf54}, 9862 (1996).

\bibitem{Starykh} O.A. Starykh and L. Balents, Phys. Rev. Lett. {\bf 98}, 077205 (2007).

\bibitem{White1} A. Weichselbaum and S.R. White, Phys. Rev. B 84, 245130 (2011).


\bibitem{Santoro}A. Santoro, A.D. Mighell, and C.W. Reimann, Acta Crystallogr.,
{\bf26}, 9979 (1970).



\bibitem{Hammar} P.R. Hammar, M.B. Stone, D.H. Reich, C. Broholm, P.J. Gibson, M.M. Turnbull, C.P.
Landee, and M. Oshikawa M,  Phys. Rev. B {\bf59}, 1008 (1999).

\bibitem{Stone}  M.B. Stone, D.H. Reich, C. Broholm, K. Leftmann, C. Rischel, C.P. Landee, and
M.M. Thurnbull, Phys. Rev. Lett. {\bf 91}, 037205 (2003).

\bibitem{Lancaster}T. Lancaster, S.J. Blundell, M.L. Brooks, P.J. Baker, F.L.
Pratt, J.L. Manson, C.P. Landee, and C. Baines, Phys. Rev. B {\bf73}, 020410(R) (2006).


\bibitem{spectrometer} S.A. Zvyagin, J. Krzystek, P.H.M. van Loosdrecht,
G. Dhalenne, and A. Revcolevschi, Physica B {\bf 346-347}, 1 (2004).

\bibitem{Jornet}  J. Jornet-Somoza, M. Deumal, M. A. Robb, C. P. Landee, M. M. Turnbull,
R. Feyerherm, and J. J. Novoa, Inorg. Chem., \textbf{49}, 1750 (2010).

\bibitem{McGregor}K.T. McGregor and S.G. Soos, J. Chem. Phys. {\bf64}, 2506  (1976).

\bibitem{Slichter} C.P. Slichter,  Principles of Magnetic Resonance (Berlin: Springer)1990.

\bibitem{Anderson} P.W. Anderson and P.R. Weiss, Rev. Mod. Phys. {\bf25}, 269 (1953).

\bibitem{Hoehenberg} P.C. Hoehenberg and B.I. Halperin,  Rev. Mod. Phys. {\bf49} 435 (1997).

\bibitem{Takayama} C. Yasuda, S. Todo, K. Hukushima, F. Alet, M. Keller, M. Troyer,
and H. Takayama, Phys. Rev. Lett. \textbf{94}, 217201 (2005).













\end{references}
\end{document}